\documentclass[conference]{IEEEtran}

\usepackage{cite}
\usepackage[pdftex]{graphicx}
\usepackage[cmex10]{amsmath}
\usepackage{algorithmic}
\usepackage{array}
\ifCLASSOPTIONcompsoc
  \usepackage[caption=false,font=normalsize,labelfont=sf,textfont=sf]{subfig}
\else
  \usepackage[caption=false,font=footnotesize]{subfig}
\fi
\usepackage[hidelinks]{hyperref}
\usepackage[T1]{fontenc}
\usepackage{url}

\usepackage[usenames,dvipsnames,svgnames,table]{xcolor}
\definecolor{blue}{rgb}{0,0,1}

\definecolor{dblue}{rgb}{0.0,0.0,0.5}

\definecolor{dgreen}{rgb}{0.0,0.5,0.0}

\definecolor{lightBlue}{rgb}{0.,0.5,0.5}

\begin{document}

\title{Secondary control activation analysed and predicted with explainable AI}

\IEEEoverridecommandlockouts

\author{\IEEEauthorblockN{Johannes Kruse\IEEEauthorrefmark{1}\IEEEauthorrefmark{2}\IEEEauthorrefmark{4}, Benjamin Sch\"afer\IEEEauthorrefmark{3}\IEEEauthorrefmark{5}, Dirk Witthaut\IEEEauthorrefmark{1}\IEEEauthorrefmark{2}} 
\IEEEauthorblockA{\IEEEauthorrefmark{1}Institute for Energy and Climate Research - Systems Analysis and Technology Evaluation (IEK-STE),\\Forschungszentrum J\"ulich,  52428 J\"ulich, Germany } 
\IEEEauthorblockA{\IEEEauthorrefmark{2}Institute for Theoretical Physics, University of Cologne, 50937 K\"oln, Germany}
\IEEEauthorblockA{\IEEEauthorrefmark{3}School of Mathematical Sciences, Queen Mary University of London, London E1 4NS, United Kingdom }
\IEEEauthorblockA{\IEEEauthorrefmark{5}Faculty of Science and Technology, Norwegian University of Life Sciences, 1432 Ås, Norway}
\IEEEauthorblockA{\IEEEauthorrefmark{4} Email: jo.kruse@fz-juelich.de }
\thanks{\noindent We gratefully acknowledge support from the German Federal Ministry of Education and Research (BMBF grant no. 03EK3055B) and the Helmholtz Association via the \textit{Helmholtz School for Data Science in Life, Earth and Energy} (HDS-LEE). This project has received funding from the European Union’s Horizon 2020 research and innovation programme under the Marie Sk\l{}odowska-Curie grant agreement No. 840825.}
}

\maketitle

\begin{abstract}
The transition to a renewable energy system poses challenges for power grid operation and stability. Secondary control is key in restoring the power system  to its reference following a disturbance. Underestimating the necessary control capacity may require emergency measures, such as load shedding. Hence, a solid understanding of the emerging risks and the driving factors of control is needed.
In this contribution, we establish an explainable machine learning model for the activation of secondary control power in Germany.  Training gradient boosted trees, we obtain an accurate description of control activation. Using SHapely Additive exPlanation (SHAP) values, we investigate the dependency between control activation and external features such as the generation mix, forecasting errors, and electricity market data. Thereby, our analysis reveals drivers that lead to high reserve requirements in the German power system. Our transparent approach, utilizing open data and making machine learning models interpretable, opens new scientific discovery avenues.
\end{abstract}

\begin{IEEEkeywords}
power grid, frequency, control, data-driven, explainable AI, machine learning 
\end{IEEEkeywords}


\section{Introduction}

Balancing and control is central for the stable operation of power systems. Secondary control is one of three measures that are typically installed to enforce the balance between power supply and demand \cite{machowskiPowerSystemDynamics2008}. While primary control  acts within a few seconds after a disturbance and stabilises the frequency, secondary control activates fully after a few minutes and restores the frequency back to its reference value. Secondary control, also known as automatic Frequency Restoration Reserve (aFRR) in Continental Europe, activates automatically according to the local power mismatch of the control area. Meanwhile, a lack of control reserves requires costly emergency measures such as load shedding. For an appropriate reserve sizing and optimal control design we thus need a precise modelling and a good understanding of the required aFRR volumes. Furthermore, predicting future aFRR volumes can be helpful for trading and bidding strategies.


Data-driven models have already proven to be excellent candidates for modelling and predicting aFRR. In the past years, power system data has become increasingly publicly available, thus enabling transparent data-driven analysis and prediction \cite{hirthENTSOETransparencyPlatform2018,morrisonEnergySystemModeling2018}. Koch et al.\ have used multiple regression and data analysis to disentangle the German paradox of increasing renewable penetration and decreasing imbalance volumes \cite{kochShorttermElectricityTrading2019}. Similarly, ref. \cite{weissbachImpactCurrentMarket2018} applies a data-driven analysis to examine the impact of 15 min intra-day trading on imbalances and control volumes in Germany. Apart from ex-post analysis, data-driven prediction methods have been developed for the aFRR market to allow for optimal bidding strategies \cite{mertenAutomaticFrequencyRestoration2020}. From the perspective of Transmission system operators (TSOs), data-driven prediction methods have been used to optimise the dimensioning of aFRR capacities based on historic data \cite{jost2015dynamic}. 
While simple probabilistic methods use parametric models to estimate the probabilities of power imbalances, more advanced methods apply Machine Learning to predict system imbalances from external features \cite{de2019dynamic}. In this context, Artificial Neural Networks \cite{jostDynamicSizingAutomatic2016}, (non-parametric) kernel density estimation and k-means clustering \cite{bucksteegImpactsDynamicProbabilistic2016} and LASSO \cite{esslMachineLearningAnalysis2017} have been applied to predict aFRR volumes. However, complex Machine learning models are often hard to interpret and their black-box character impedes their application in security-relevant areas such as power system control \cite{ahmadArtificialIntelligenceSustainable2021a}. 

Here, we present an explainable ML model for aFRR to enhance the value of Machine Learning for power system operation and control.  Using publicly available data \cite{ENTSOETransparencyPlatform, regelleistungWebsite}, we build an ML model for the ex-post analysis of historic data as well as a day-ahead predictor. We interpret the model with SHapely Additive exPlanation (SHAP) values \cite{lundbergLocalExplanationsGlobal2020a}, which enable rich explanations of ex-post models as well as transparent day-ahead predictors. As our case study, we focus on the German aFRR and demonstrate how loss functions and data sets have to be adapted for either use case. Our publicly available data set \cite{kruseDataSetPredictionAFRR} comprises two years of 15min resolved data with 85 features from the German and European power system as well as the corresponding aFRR volumes. 

We start in Section~\ref{sec:methods} by describing our data collection and pre-processing procedure as well as the Machine Learning model. In Section~\ref{sec:expost_analysis}, we apply our model for an ex-post analysis of dependencies and important features in the German aFRR system. In Section~\ref{sec:dayahead_prediction}, we switch to the day-ahead prediction of aFRR volumes and demonstrate the impact of different loss functions and feature sets, before closing with a discussion in Section~\ref{sec:discussion}.

\begin{figure*}[tb]
\centering
\includegraphics[width=2\columnwidth]{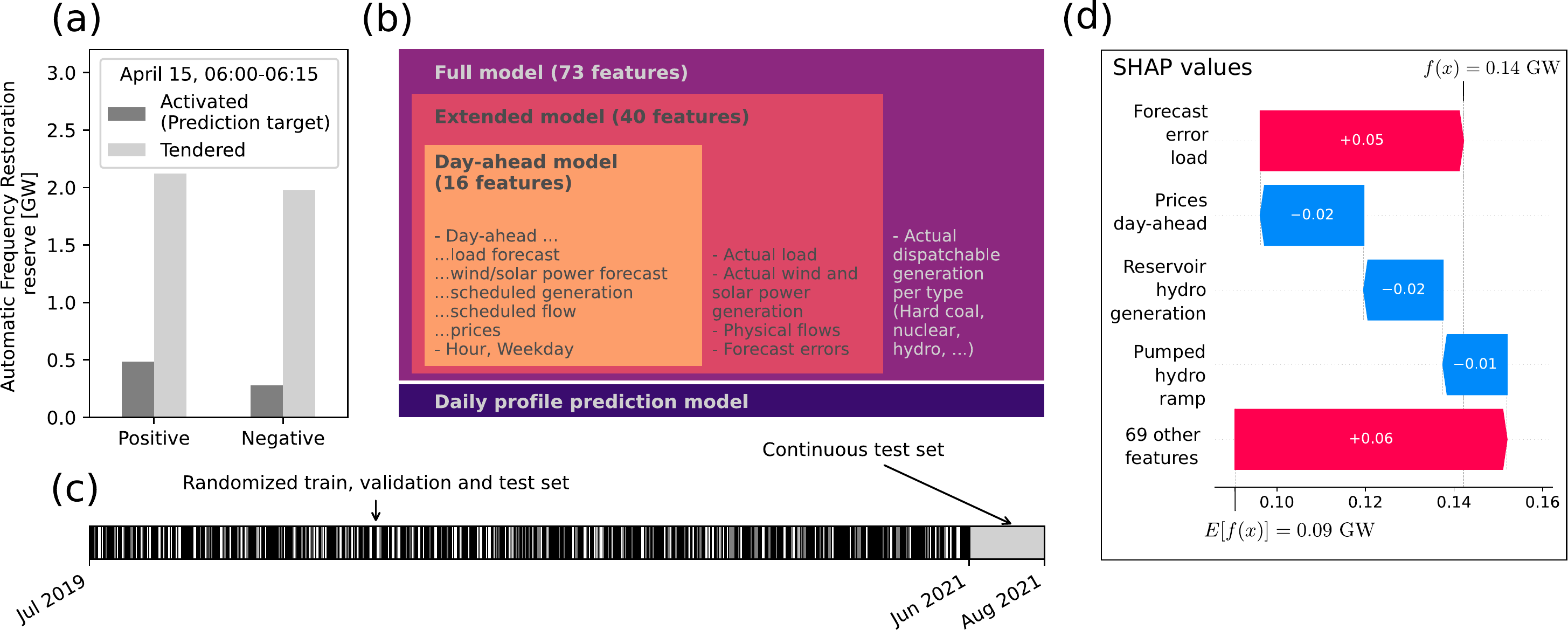}
\caption{
\label{figmodel-over}
Building an explainable Machine Learning model for activated secondary control (aFRR). (a): Our model predicts the activated amount of positive and negative aFRR in Germany with a 15 min resolution. We downloaded activated power and tendered power demands from the German TSOs \cite{regelleistungWebsite}, which is depicted here for one time step in 2021. (b) We considered four different models. A daily profile, only dependent on the historic aFRR, and three models utilising data from the ENTSO-E transparency platform \cite{ENTSOETransparencyPlatform}:  a prediction-oriented model using only day-ahead values, an extended model including actual load, wind/solar generation, physical flows and thereby forecast errors and a full model that also comprises actual dispatchable generation. (c): We trained and tested these models on randomised train, validation and test sets with data from July 2019 to May 2021. The last two month of the data set served as a continuous test set for time series forecasting. (d): To explain our model predictions, we used SHAP values, which quantify the impact of each feature on the model prediction $f(x)$ relative to the expected prediction $E[f(x)]$ (the base value). The figure refers to the prediction of Positive aFRR for the time step used in panel (a).  
}
\end{figure*}

\section{Methods}
\label{sec:methods}

\subsection{Frequency restoration reserve: Markets and data}
\label{sec:afrr_data}
The design of balancing markets determines the procurement and activation of control power. In Germany, the TSOs procure their aFRR demand through an anonymous auction \cite{mertenAutomaticFrequencyRestoration2020, regelleistungWebsite}. First, suppliers of aFRR have to fulfil pre-qualification criteria such as a minimum activation speed to participate in the auction. As of September 2020, hydro- and gas-driven power plants represent the largest part (75\%) of the prequalified aFRR capacity \cite{prequalifiedCapacityGermany}. Second, the TSOs tender a demand for the required reserve capacity and energy, which is done  for all four German TSOs together (grid control cooperation). Then, prequalified parties can sell reserve capacity (in MW) on the capacity market, which they must hold available. In addition, a supplier can omit this step and directly bid at the energy market (in MWh). The energy market determines the suppliers that actually deliver the balancing energy.

The activation of aFRR depends on the local power imbalance. The TSOs activate aFRR according to the imbalance of the control area. The activation is typically proportional to the integrated imbalance, which is the case for a standard PI-controller \cite{machowskiPowerSystemDynamics2008}. Suppliers with bids on the energy market are activated successively starting with the lowest price until the demand is met. However, the International Grid Control Cooperation (IGCC) further changes the activated control volume \cite{regelleistungWebsite}. The IGCC avoids the activation of counter-acting aFRR in different countries through an imbalance netting among the 17 operational IGCC member states in Europe \cite{entso-eENTSOEBalancingReport}. 

Both the procurement and activation of aFRR exhibit fixed time scales and deadlines. The TSOs tender capacity demands one week ahead and the capacity market closes one day ahead of delivery, while the energy market closes one hour ahead. On both markets, aFRR is sold for 4h blocks separately for negative (downward) and positive (upward) regulation. These time scales also determine the resolution of the available aFRR data.

For our prediction study, we used publicly available aFRR data from July 2019 to July 2021 \cite{regelleistungWebsite}. We downloaded tendered capacity demands, which come with a 4h resolution, and activated aFRR with a resolution of 15min (Fig.~\ref{figmodel-over}a). We used the Germany-wide activated aFRR (in GW) as the target of our prediction, while the tendered demands only serve as a benchmark. We note, that the market design during our period of investigation changed: The energy market was only introduced in November 2020. Before, both the capacity price and the energy price were submitted to the day-ahead capacity market and energy bids with shorter lead times were not possible. 

\subsection{Input features and prediction models}
\label{sec:features}
As inputs for our prediction model, we used publicly available power system features from the ENTSO-E Transparency platform \cite{ENTSOETransparencyPlatform}. The feature preparation included data collection, aggregation, upsampling and feature engineering. 

Following \cite{kruseRevealingRisks2021}, we collected six feature types for Germany from the ENTSO-E Transparency platform: Day-ahead forecast data for wind and solar power, load forecasts, day-ahead scheduled generation, day-ahead prices, actual generation per type and actual load. In addition, we included pumped hydro consumption and cross-border power flows in this study. The flows comprise day-ahead and total commercial exchanges and physical flows. For each flow feature, the in and out flows between Germany and its neighbours were directly aggregated into one import-export balance reflecting the total (positive or negative) flow into Germany. 

To model the impact of imbalance netting (IGCC), we aggregated the features across the other IGCC member states (excluding Germany). We only included day-ahead forecast data, since the actual imbalance netting between the countries should be reflected in the actual cross-border flows. The aggregation follows the procedure in ref. \cite{kruseRevealingRisks2021}. 

In line with the prediction target, we use features with 15min resolution. The German data mostly exhibits the required 15min resolution, except from day-ahead prices, scheduled generation and flow variables. These have an hourly resolution, so that 4 hourly steps were padded with the same value. The same procedure is applied to IGCC features that often come with only 1h resolution. Day-ahead load and renewable forecasts are treated differently; here we used a linear interpolation for upsampling for the continuous nature of the variable. 

To enhance interpretability, we finally added engineered features to our input data. For each feature, we constructed ramps (gradients from time $t-1$ to $t$) and day-ahead forecast errors. Positive forecast errors indicate an overestimation of the actual values by the day-ahead forecast. The flows yield two forecast errors, the day-ahead error between commercial exchanges and physical flows, as well as the unscheduled flows (the difference between total exchanges and physical flows).

From this data, we constructed four models containing different feature sets (Fig.~\ref{figmodel-over}b): The day-ahead model contains only day-ahead available data, the extended model also includes actual load, renewable generation, physical flows and thereby forecast  errors and the full model that also comprises actual conventional generation. Conventional generators such as hydro or nuclear power participate in frequency control \cite{prequalifiedCapacityGermany}, which opens the question whether we can predict the activated control \textit{without} including the actual output of participating generation types. The fourth model, the daily profile, predicts the daily mean evolution of the activated aFRR based on historic aFRR data, without using additional features. We use a 5th model in Section~\ref{sec:dayahead_prediction}, which is a variation of the day-ahead model. It additionally contains the day-ahead features (day-ahead load, wind and solar forecasts, ...) aggregated over the other IGCC member states.


\subsection{Model training, evaluation and interpretation}
\label{sec:model}
We used Gradient Boosted Trees (GBTs) to predict the activated aFRR from power system features. GBTs offer complex non-linear models and perform inherent feature selection \cite{hastieElementsStatisticalLearning2016}, which is beneficial for the case of strongly correlated time series features in the power system \cite{kruseRevealingRisks2021}. Moreover, tree-based methods are highly interpretable and offer efficient ways to compute model explanations  \cite{lundbergLocalExplanationsGlobal2020a}. We used the LightGBM implementation of GBTs, which enables a particularly fast way of model training \cite{keLightGBMHighlyEfficient2017}. 


To train our model, we split the data set into train, test and validation set (Fig.~\ref{figmodel-over}c). First, we set aside the last two months of the data set as a continuous test set for time series forecasting. Then, we randomly split the remaining part into a train set (64\%), a validation set (16\%) and a test set (20\%). We optimised the hyper-parameters of our LightGBM model via grid search and 5-fold cross validation on the train set, while performing early stopping of the boosting rounds on a validation set. Finally, we retrained the model with optimal hyper-parameters on the union of the train and validation set. This model was used to evaluate the performance on the randomised or the continuous test set. Note that we used different loss functions for training and different evaluation metrics, which we will specify depending on the use case.

We interpreted the trained model with SHapely Additive exPlanation (SHAP) values \cite{lundbergLocalExplanationsGlobal2020a}. SHAP values quantify the (positive or negative) impact of each feature on an individual model prediction relative to a base value. They avoid inconsistencies present in other feature attribution methods \cite{lundbergConsistentIndividualizedFeature2019} and fulfil certain optimal properties \cite{lundbergLocalExplanationsGlobal2020a}. For example, their \textit{local accuracy} guarantees that SHAP values sum up to the model prediction. Fig.~\ref{figmodel-over}d depicts a sample prediction, where "Forecast error load" has a positive  and "Prices day-ahead" has a negative impact on the prediction. Adding up all feature contributions and the base value (the expected prediction) yields the model output $f(x)$.  

\subsection{Data and code availability}
Our data set, including the prediction targets and features, as well as the results of our hyper-parameter optimisation are available on Zenodo \cite{kruseDataSetPredictionAFRR}. The code, including the details on our feature pre-processing and the hyper-parameter optimisation, is accessible on GitHub \cite{afrr_prediction_repo}. 

\section{Ex-post analysis of aFRR operation}
\label{sec:expost_analysis}
\begin{figure}[tb]
\includegraphics[width=\columnwidth]{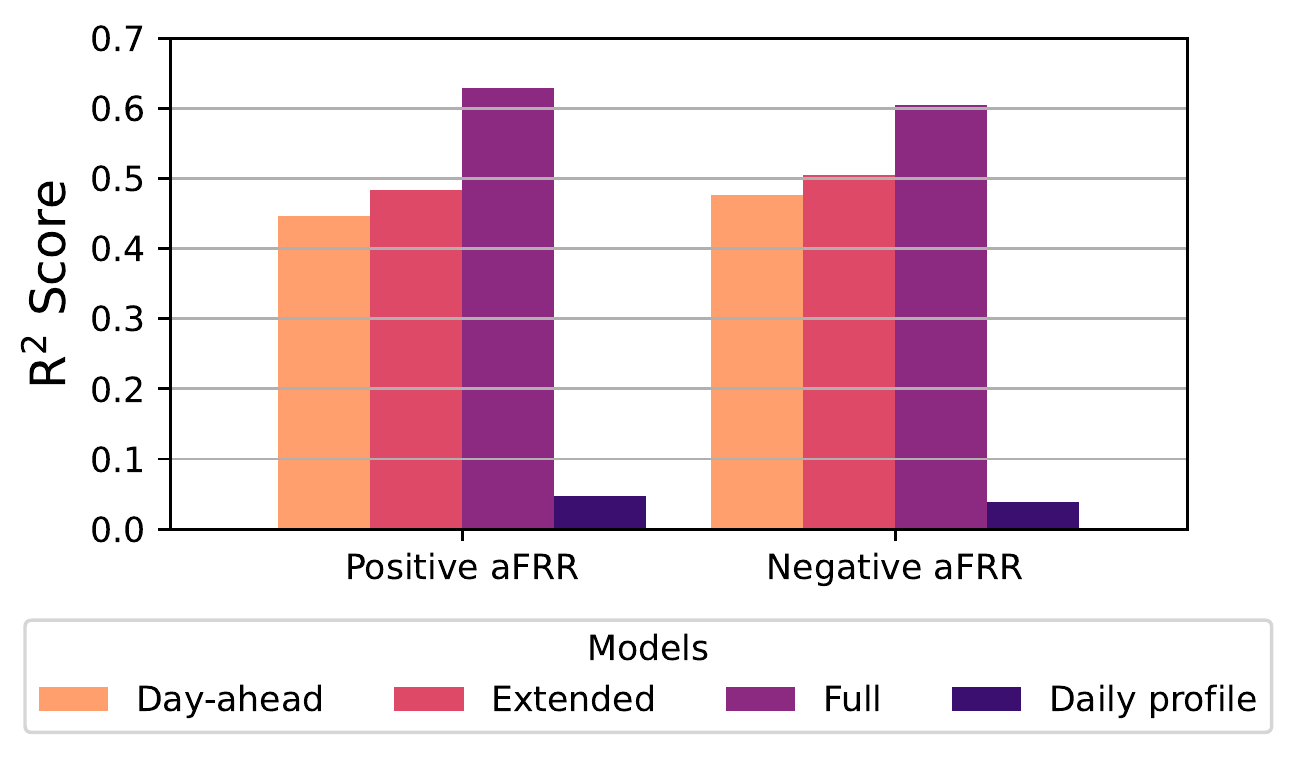}
\caption{
\label{fig:expost-r2-score}
Prediction performance varies strongly with the model type and feature set. For our ex-post analysis, we trained our model with an L2-loss and used the randomised train and test sets to predict positive (left) and negative (right) activated aFRR. In terms of the R$^2$-score, the full model including actual hydro power generation performed best which is likely a manifestation of reversed causality as hydro power supplies most of the aFRR in Germany. Daily patterns in the aFRR activation are very weak such that the daily profile performed badly. 
}
\end{figure}

\begin{figure*}[tb]
\centering
\includegraphics[width=\textwidth]{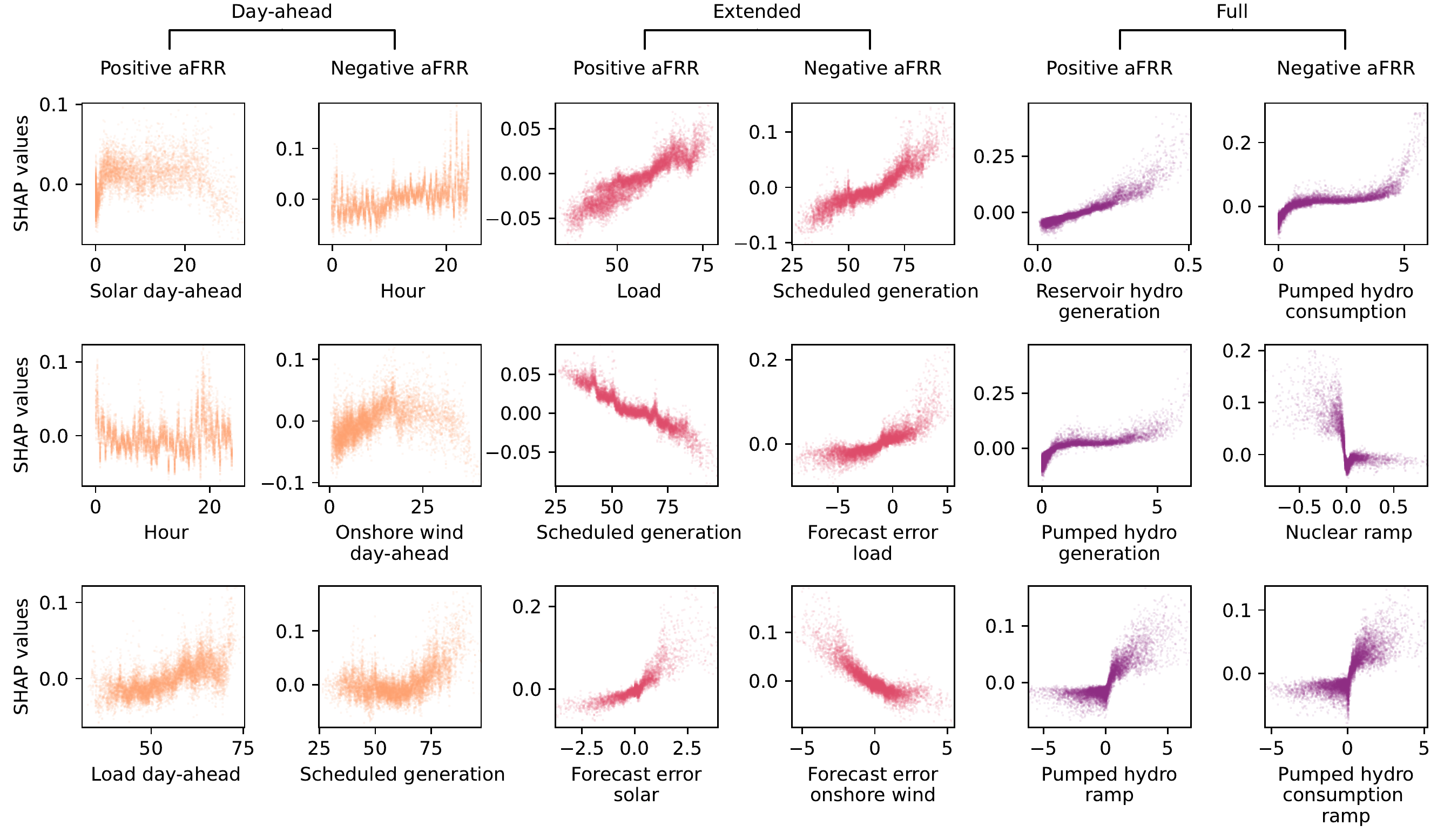}
\caption{
\label{fig:expost-importances}
Partial dependency plots can reveal important drivers of activated aFRR depending on the model type. Dependency plots depict the relation between SHAP values and the feature value, thus characterising the dependency of the model output on the feature. We show the three most important features for the day-ahead (left), extended (center) and full model (right) with decreasing mean absolute SHAP value (from top to bottom). The day-ahead and extended models did not include actual generation that supplies aFRR. Thus, the most important dependency plots for these models likely reflect causal drivers such as forecast errors and mismatches between load and scheduled generation. In contrast, the full model fitted relations between hydro power and aFRR activation, which probably reflect a reverse causation and must be interpreted differently in the ex-post analysis. 
}
\end{figure*}

\begin{figure*}[tb]
\centering
\includegraphics[width=0.92\textwidth]{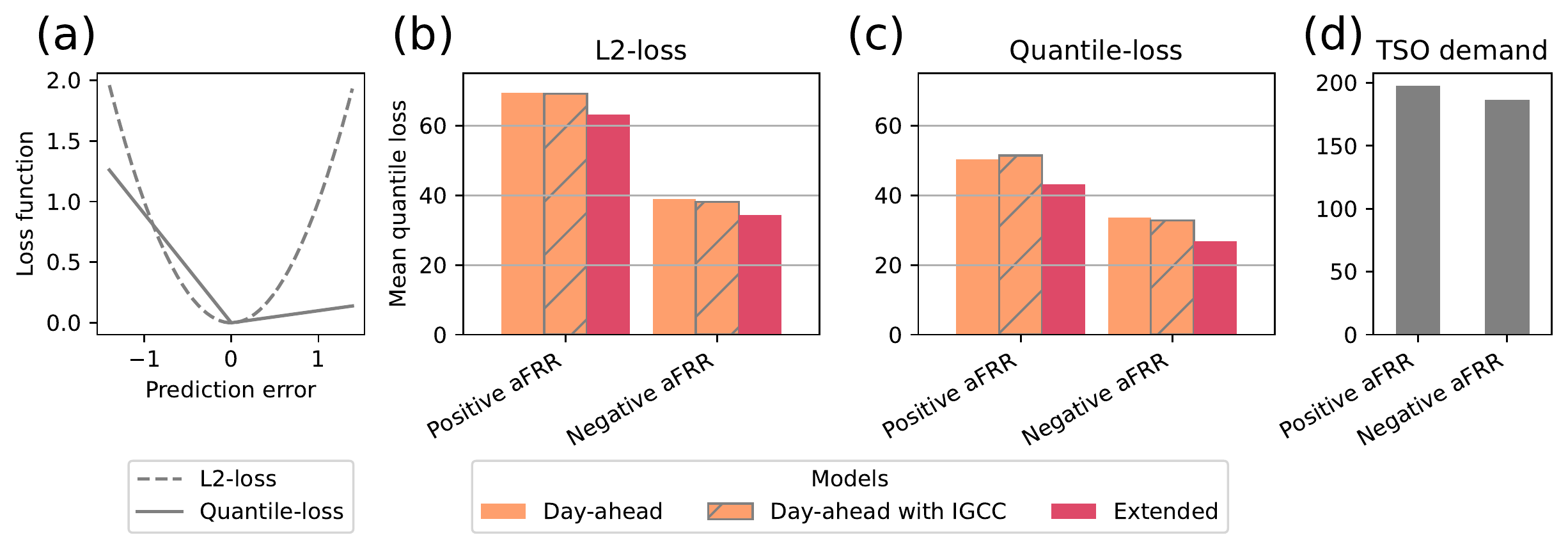}
\caption{
\label{fig:loss-function}
Loss functions have to fit the use case of the model. (a): For the day-ahead prediction of activated aFRR, we trained our model both with the L2-loss and the Quantile-loss (90 \% quantile) and evaluated its performance on the continuous test set (cf. Fig.~\ref{figmodel-over}c). In this figure, a positive prediction error indicates an overestimation of the true value by the prediction. (b) and (c): The mean quantile loss (90 \% quantile) represents the weighted costs of over- and underestimating the aFRR volume, where an underestimation is much more costly than procuring too much aFRR. The Quantile-loss model performed best and further minimized these costs compared to the L2-loss model. While the L2-loss suits best for explaining the aFRR trajectory ex-post, the Quantile-loss better fits the task to predict required aFRR volumes day-ahead. (d): Tendered aFRR demands from the Transmission System Operators (TSOs) strongly overestimated the activated aFRR in most cases thus yielding a high mean quantile loss. 
}
\end{figure*}

We first consider the application of our  ML model for the ex-post analysis of the aFRR system. Model explainability is essential for any analysis, and we resorted to SHAP values for this task (cf. Section~\ref{sec:model}). For analysis purposes, we aimed at reproducing the system trajectory as well as possible and thus chose a standard L2-loss function (cf. Fig.~\ref{fig:loss-function}) and evaluated the performance of the model by the $R^2$ score. For model training, evaluation and interpretation we used the randomised train and test set (cf. Fig.~\ref{figmodel-over}c).

We found that the tree-based model reproduces the true system trajectory with an $R^2$ score in the range 0.45-0.63 depending on the feature used in training set and analysis (Fig.~\ref{fig:expost-r2-score}). Hence, we conclude that the ML model accounts for roughly half of the variability.

Taking into account the full set of features provides the best $R^2$ score as expected, but it does not necessarily provide the deepest insight into the drivers of the aFRR system. This was revealed by the SHAP framework, which quantifies both the relation of individual predictions and feature values as well as the global feature importance. Inspecting the results for the full feature set in figure~\ref{fig:expost-importances}, we found that the most important features are given by actual generation or generation ramps of hydro power plants and the ramp of nuclear power generation in the case of negative aFRR. The hydro power plants can be switched and controlled rather rapidly and thus provide a major share of aFRR power. In fact, they make up for the largest amount of prequalified aFRR capacity \cite{prequalifiedCapacityGermany}.
Hence, it is likely that we here observed a case of reverse causation: The application of aFRR caused a strong activity of the respective hydro power plants and \emph{not} vice versa. Hence, the model with the full feature set rather explained how aFRR is provided and not why. For instance, the model suggested that positive aFRR is predominantly provided by increasing hydro generation while negative aFRR predominantly provided by increasing pumped hydro consumption. One might extend the analysis to further infer from data which generation types actually contribute to aFRR provision. 

In the case of negative aFRR, nuclear power ramps were among the most important features. Nuclear power plants can in principle provide aFRR, but they account for less than 2\% of the total prequalified capacity in Germany \cite{prequalifiedCapacityGermany}. Hence, we conclude that the dependency of aFRR activation and nuclear ramps likely reflects a causal relation. Nuclear power plants typically ramp slowly and continuously, which can lead to imbalances of power generation and load. In fact, a previous study has shown a strong dependency between nuclear ramps and long-lasting deviations of the grid frequency \cite{kruseRevealingRisks2021}, which then results in the activation of aFRR. 

Restricting the model to the extended feature set provided a different picture. The $R^2$ score decreases to 0.50 (negative aFRR) and 0.48 (positive aFRR), respectively. In this case, the most important features were given by the scheduled generation, the load and forecasting error of load and renewable generation (Fig.~\ref{fig:expost-importances}). 
The high importance of forecasting errors is absolutely consistent with our expectation on the causal interrelation of the aFRR systems. Forecasting errors generally lead to an imbalance of generation and load causing a deviation of the grid frequency from its reference value. The control system will thus demand the activation of aFRR power.

The high feature importance of the scheduled generation and the total load reflects the general requirements for secondary control power: If the scheduled generation is already high, it is more likely that it has to be reduced via negative aFRR than it has to be increased via positive aFRR. Similarly, if the scheduled generation is already low, it is more likely to be increased via positive aFRR than it has to reduced via positive aFRR. Hence, the SHAP values for positive aFRR increased with the scheduled generation, while the SHAP values for negative aFRR decreased with the scheduled generation. The reverse relation was found for the load. 
The large impact of actual load and scheduled generation (with reverse signs) suggests that the difference between actual load and scheduled generation per 15min interval is very relevant for the aFRR. This intuitively makes sense as the aFRR activation depends on the area control error, see also \cite{weissbachImpactCurrentMarket2018}.

Remarkably, we did not observe a high feature importance of the ramps of load and the dispatchable generation, which are central for the understanding of deterministic frequency deviations (DFDs), in particular the Rate of Change of Frequency (RoCoF)  \cite{kruseRevealingRisks2021}. These DFDs are caused by short-term imbalances due to different adaption/ramping behaviour of generation and load \cite{weissbachHighFrequencyDeviations2009}. The low importance of ramps in the aFRR model suggests that DFDs are mostly compensated by primary control. The low importance of DFDs for the aFRR activation is consistent with other studies that show a strong decrease in deterministic aFRR peaks between 2012 and 2018 due to the introduction of 15min intra-day trading in December 2011 \cite{weissbachImpactCurrentMarket2018, kochShorttermElectricityTrading2019}.


In the day-ahead model, the most important features were the total generation (or similarly load), as well as volatile renewable forecast and the hour of the day. Dependency plots showed a strong vertical dispersion, such that the dependencies were less clear than in the other models. This observation, as well as the high importance of the hour, showed that more specific information is missing in this model. 
Notably, load and generation were both used heavily by the day-ahead and the extended models with similar but different partial dependencies: high generation led to larger prediction in negative aFRR, while high load led to high predictions in positive aFRR.

Finally, we remark that a good analysis model is not necessarily a good prediction model. In the current framework this becomes apparent if we do not choose a randomized test set, but a continuous test set in the end of the available data interval, i.e., replace an interpolation by an extrapolation task. In this case the $R^2$ score dropped dramatically to values below 0.38 (full model), 0.12 (extended model) and 0 (day-ahead model), respectively (plots not shown). This might be related to the fact that the entire aFRR system evolves quite strongly during time, for instance regulatory framework has been adapted repeatedly during the analysis period (cf. Section~\ref{sec:afrr_data}).


\section{Day-ahead prediction of activated aFRR}
\label{sec:dayahead_prediction}

The L2-loss used in the analysis part was well-suited to predict the actual aFRR on average but often underestimated the necessary control. Indeed, the L2-loss, treats predictions underestimating control identical to those overestimating it, while in reality a shortage of control is much more costly: If the necessary control exceeds the estimated control, i.e., the control available in back-up generators, the frequency cannot be restored back to its original set point, making it vulnerable in case of further disturbances and thereby increasing the risks of generator disconnections and load shedding \cite{machowskiPowerSystemDynamics2008}. 

Hence, moving from a pure ex-post analysis towards a predictive model, we introduced a new loss function, namely the Quantile-loss, see Fig.~\ref{fig:loss-function}. Similar to the L2-loss it penalises predictions more the further they are away from the actual value. In contrast to the L2-loss, the Quantile-loss penalises asymmetrically: Overestimating costs is punished much less than underestimating costs (Fig.~\ref{fig:loss-function}a). In particular, we employed the 90\% quantile, i.e., we expect to overestimate the target value in 90\% of the cases by penalising underestimation nine times more than overestimation. Consistently, we did not compute the $R^2$ score when evaluating model performance but the mean quantile loss, which can be interpreted as the weighted average cost of false prediction and procurement of aFRR capacity. Any underestimation of control needs causes larger costs than any overestimation. Here, we consider the 90\% quantile corresponding to weight factors of $10:1$, but the model can be readily adjusted for the actual costs of a TSO.
In contrast to the previously used $R^2$-score, a model performs better when its mean quantile loss is low.

\begin{figure*}[tb]
\centering
\includegraphics[width=0.9\textwidth]{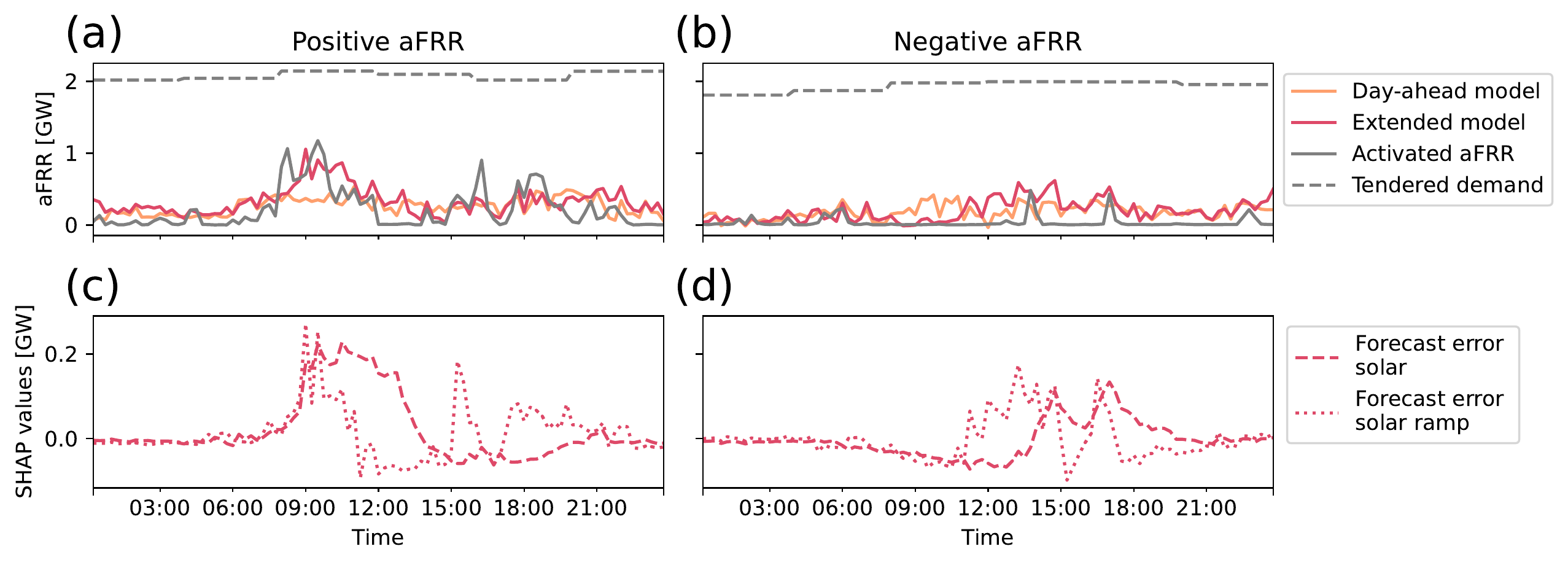}
\caption{
\label{fig:trajectoires_explained}
SHAP values identify renewable forecast errors as an error source for day-ahead aFRR prediction. (a) and (b): The . depict the prediction of the Quantile-loss models for July 14, 2021, as part of the continuous test set (cf. Fig.~\ref{figmodel-over}c). While the day-ahead model already remained above activated aFRR volumes for many time steps, it underestimated the volumes in certain periods, where the extended model performed better. (c) and (d): The SHAP values represent the impact of solar forecast errors on the prediction of the extended model (trained with Quantile-loss). For example between 9:00 and 10:00, the extended model better estimated the positive aFRR mostly due to the large impact of solar forecast errors in the model, thus pinpointing the source of prediction errors in the day-ahead model.   
}
\end{figure*}

Indeed, when comparing models trained on L2-loss and Quantile-loss, we noted a substantial improvement by introducing the adequate loss function (Figs.~\ref{fig:loss-function}b and \ref{fig:loss-function}c). Within each loss function, moving from the day-ahead to the extended model again yielded a further improvement, consistent with the results from Section~\ref{sec:expost_analysis}. Furthermore, we considered an expansion of the day-ahead model where we included the  International Grid Control Cooperation (IGCC) as a new feature, see Section~\ref{sec:features} for details. Including the IGCC altered the prediction quality only slightly, regardless of the chosen loss function, and hence we did not investigate it in detail in the following. Finally, the TSO (tendered) demand systematically overestimated the necessary control by far (see also Fig.~\ref{fig:trajectoires_explained}) and hence led to the highest costs according to the loss considered here (Fig.~\ref{fig:loss-function}d).

The differences between day-ahead, extended and tendered demand became even more clear when visualising aFRR time series (Fig.~\ref{fig:trajectoires_explained}a and \ref{fig:trajectoires_explained}b): The tendered capacity demand was almost constant and always overestimated the actual demand substantially. The true time series showed pronounced peaks, which were not fully reproduced by the day-ahead model. Meanwhile, the extended model more closely resembled the true time series. 
Using SHAP, we were able to investigate why the predictions of the day-ahead and the extended model differed and thereby identify forecast errors in solar generation and ramps as the main reasons (Figs.~\ref{fig:trajectoires_explained}c and \ref{fig:trajectoires_explained}d).

\begin{figure*}[tb]
\centering
\includegraphics[width=0.9\textwidth]{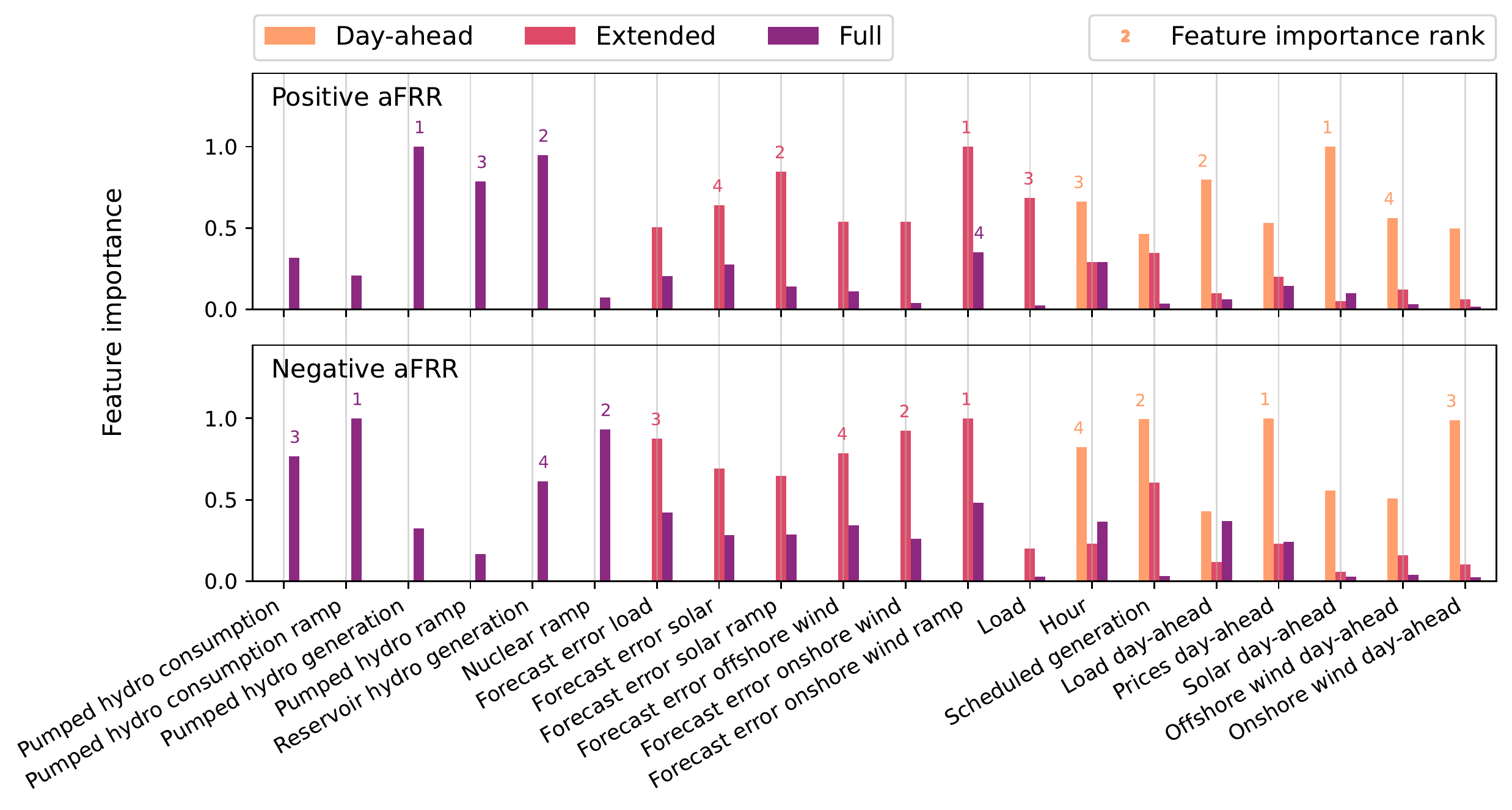}
\caption{
\label{fig:all-importances}
An overview of feature importances reflects consistent overlaps as well as differences between the model types. We display the mean absolute SHAP value as a measure for the feature importance for all three model types trained with the Quantile-loss (90\% quantile). To be able to compare with Fig.~\ref{fig:expost-importances}, we evaluated the SHAP values on the randomised test set (cf. Fig.~\ref{figmodel-over}c). The figure shows the union of the four most important features across all model types and targets and the numbers on top of the bars indicate the rank of the four most important features. From left to right, the models become more restrictive in the feature set thus resulting in differences between the most important features. However, (day-ahead) load as well as features related to wind and solar power were consistently important in all models.
}
\end{figure*}

To obtain a comprehensive comparison between day-ahead, extended and full model using the Quantile-loss, we compared their most important features in Fig.~\ref{fig:all-importances}. Based on the design of the models, only the full model could use specific actual generation information, e.g., on hydro or nuclear power, while forecast errors were available to both extended and full model and the day-ahead model was restricted to estimates.
Intriguingly, each model used some of its unique features extensively and hence they typically had a high rank, yielding results consistent with the L2-loss results discussed in Fig.~\ref{fig:expost-importances}.
Although different features were available, we still noticed some overlap: For positive aFRR both the extended and the day-ahead model utilised load (or load-day-ahead) as a top feature, while negative aFRR was predicted also based on forecast errors of load and scheduled generation respectively. This indicates that the absolute value of the load (or the necessary generation) is important for predicting aFRR usage. 
Furthermore, the extended model heavily used forecast errors of renewable generation, while the day-ahead model relied on the day-ahead renewable forecasts. This suggests that our day-ahead model tried to estimate potential forecast errors from the day-ahead generation forecasts since forecast errors appeared as the actual drivers of control activation, indicated by the increased performance of the extended model. Ideally, forecast errors would be as small as possible, thereby reducing the need for aFRR. If this is not possible, including uncertainties in the forecasts might give better estimates of necessary aFRR. 


\section{Discussion}
\label{sec:discussion}
Concluding, we have demonstrated how boosted trees and SHapley Additive exPlanations (SHAP) offer versatile tools for investigating secondary frequency control both for ex-post interpretation and also when forecasting trajectories.

Including all available features in the ex-post analysis yielded the most accurate description, both for negative and positive aFRR. Interestingly, the aFRR behaved very differently from deterministic frequency deviations (DFDs). DFDs are mostly driven by ramps \cite{kruseExploring2021} and are already well-described by daily profiles \cite{weissbachHighFrequencyDeviations2009,weissbachImpactCurrentMarket2018}. Contrary, the aFRR depended much more on the mismatch between actual load and scheduled generation due to forecasting errors and not as much on ramps, thus not showing a pronounced daily profile. 

The day-ahead forecasts of the necessary aFRR mostly relied on day-ahead estimations for the volatile renewable generation and also on the total load and generation. Using SHAP, we did not only obtain a most open and interpretable model but also had the opportunity to identify the cause for mispredictions, e.g., pinpointing them to large solar generation forecast errors.

Several important lessons are to be learnt when applying machine learning to power system analysis: 
Firstly, good performance and interpretability of machine learning models can be achieved by combining complex models with ex-post interpretations, such as boosted trees combined with SHAP used here, or right-off starting with white-box models, i.e., choosing techniques that are inherently interpretable \cite{barredoarrietaExplainableArtificialIntelligence2020}. Hence, they allow to reduce the usage of black-box machine learning models, which pose severe security concerns \cite{ahmadArtificialIntelligenceSustainable2021a}.
Secondly, the selection of input features is critical when answering research questions via machine learning: Using ex-post analysis of all available generation data, we observed hydro power as critical. But this is likely a reverse causality: Generation in hydro power plants does not raise the need for secondary control but the necessary secondary control is provided by hydro power plants. Furthermore, when excluding features, in our case moving from the full towards the day-ahead model, different features will be used for similar predictions. Therefore, before deeming a feature essential based on a single feature set, regularisation methods \cite{hastieElementsStatisticalLearning2016} should be considered and different feature sets used to train a model.
Thirdly, loss functions are critical to tune the model towards desired performance. In the case of control power, underestimating the power is much more costly than overestimating it and this has to be reflected in the loss function. Ideally, loss functions are directly related to the actual costs of false predictions in the system under investigation. While a square loss is adequate as an ex-post analysis tool, a Quantile-loss is more appropriate for predictions.

Frequency restoration capacities have previously been estimated using machine learning \cite{jostDynamicSizingAutomatic2016, bucksteegImpactsDynamicProbabilistic2016, esslMachineLearningAnalysis2017}, including the application of quantile loss functions. Compared to these earlier studies, we make a clear comparison between forecasting and ex-post analysis and use the same methods for both tasks, namely boosted trees analysed via SHAP values. Thereby, we emphasise the interpretability of a machine learning approach instead of only optimising performance. By interpreting analysis results, we obtain insights into the system itself and thereby generate value beyond individual algorithms.

Our study of day-ahead aFRR forecasting using different loss functions is a starting point to develop appropriate predictors for aFRR demand. Such a prediction could then be used to optimise the capacity procurement for aFRR day-ahead thus saving costs and freeing flexible capacity for other usage.
To move towards such as general predictor, the presented analysis can be further extended, e.g., by incorporating load and generation data from other countries within the same synchronous area or investigating the aFRR in other synchronous areas. Furthermore, SHAPs offer further analysis tools, such as interaction analysis \cite{kruseRevealingRisks2021} not used here in detail, which could prove useful to disentangle the influence of individual features on aFRR needs.

\bibliographystyle{IEEEtran}
\bibliography{references.bib}

\end{document}